\documentclass[aip,apl,twocolumn]{revtex4}
\usepackage{graphicx}

\begin{document}
\title{Spin-Orbit Torque Induced Spike-Timing Dependent Plasticity}
\author{Abhronil Sengupta}
\author{Zubair Al Azim}
\author{Xuanyao Fong}
\author{Kaushik Roy}
\affiliation{School of Electrical \& Computer Engineering, Purdue University, West Lafayette, IN, 47906, USA}
\date{December 2014}

\begin{abstract}
{\small
Nanoelectronic devices that mimic the functionality of synapses are a crucial requirement for performing cortical simulations of the brain. In this work we propose a ferromagnet-heavy metal heterostructure that employs spin-orbit torque to implement Spike-Timing Dependent Plasticity. The proposed device offers the advantage of decoupled spike transmission and programming current paths, thereby leading to reliable operation during online learning. Possible arrangement of such devices in a crosspoint architecture can pave the way for ultra-dense neural networks. Simulation studies indicate that the device has the potential of achieving pico-Joule level energy consumption (maximum $2 pJ$ per synaptic event) which is comparable to the energy consumption for synaptic events in biological synapses.}
\end{abstract}

\maketitle

Large scale cortical brain simulations on present day supercomputers, based on Von-Neumann model of computation, have proved highly inefficient with respect to the ultra-high density and  energy efficient processing capability of the human brain. For instance, the IBM Blue Gene supercomputer consumed 1.4$MW$ of power to simulate 5 seconds of brain activity of a cat ~\cite{ibm}. On the contrary, the human brain consumes power of the order of a few Watts. In order to harness the remarkable efficacy of the human brain in cognition and perception related tasks, the field of neuromorphic computing attempts to develop non Von-Neumann computing models inspired by the functionality of the basic building blocks, i.e. neurons and synapses in the biological brain.

The computational fabric of the brain consists of a highly interconnected structure where neurons are connected by junctions termed as synapses. Each synapse is characterized by a conductance and helps to transmit weighted signals in the form of spikes from the pre-neuron to the post-neuron. It is now widely accepted that synapses are the main computational element involved in learning and cognition. The theory of Hebbian Learning ~\cite{hebbian} postulates that the strength of synapses are modulated in accordance to the temporal relationship of the spiking patterns of the pre-neurons and post-neurons. In particular, Spike-Timing Dependent Plasticity (STDP) has emerged as one of the most popular approaches of Hebbian Learning ~\cite{bipoo}. According to STDP, if the pre-neuron spikes before the post-neuron, the conductance of the synapse potentiates (increases) and vice versa. The relative change in synaptic strength decreases exponentially with the timing difference between the pre-neuron and post-neuron spikes.The timing window during which such plastic synaptic learning occurs has been observed to be of the order $\sim 100 ms$.

The number of synapses also outnumber the number of neurons in the mammalian cortex by a large extent. The human brain consists of approximately $10^{11}$ neurons and $10^{15}$ synapses, thereby requiring $10^4$ synapses for every neuron. Although there have been several attempts to emulate synaptic functionality by CMOS transistors ~\cite{modha1, modha2}, the area overhead and power consumption involved is quite large due to the significant mismatch between the CMOS transistors and the underlying neuroscience mechanisms. As a result, nanoscale devices that emulate the functionality of such programmable, plastic, Hebbian synapses have become a crucial requirement for such neuromorphic computing platforms. To that end, researchers have proposed several programmable devices based on phase change materials ~\cite{pcm1,pcm2}, $Ag-Si$ memristors ~\cite{agsi}, chalcogenide memristors ~\cite{chalco} that mimic the synaptic functionality. Neuromorphic computing architectures employing such memristive devices have been also demonstrated ~\cite{neuro1, neuro2, neuro3}. However, nanoscale devices attaining the ultra-high density ($10^{11}$ synapses per $cm^{-2}$) and low energy consumption ( $\sim 1 pJ$ per synaptic event) of biological synapses have still remained elusive. In order to address some of the challenges involved in the search for an ideal \textquoteleft electronic\textquoteright \ synapse, we propose a device structure based on a ferromagnet with oppositely polarized magnetic domains separated by  a transition region termed $domain$ $wall$ ~\cite{dwm1,dwm2}. We will refer to such a ferromagnetic material as a domain-wall magnet (DWM) for the rest of this text. 

Recently spin-orbit torque generated by a HM has emerged as one of the promising mechanisms to manipulate the magnetization of a ferromagnet lying on top due to high spin injection efficiency ~\cite{she1,she2}. In particular, ~\cite{sayeef} demonstrates deterministic control of domain wall motion in a ferromagnet by orthogonal current flow in an underlying heavy metal (HM) layer in presence of an external magnetic field. In this work, we exploit this functionality to propose a device structure with decoupled spike transmission and learning current paths which is a crucial requirement for such neuromorphic computing architectures since the STDP learning event can take place at any time during the operation of the network. Spin-orbit torque generated by the underlying heavy metal (HM) in the magnetic heterostructure serves as the basic physical phenomena responsible for generating STDP. 

\begin{figure}[h]
\centering
\includegraphics[width=2.8in]{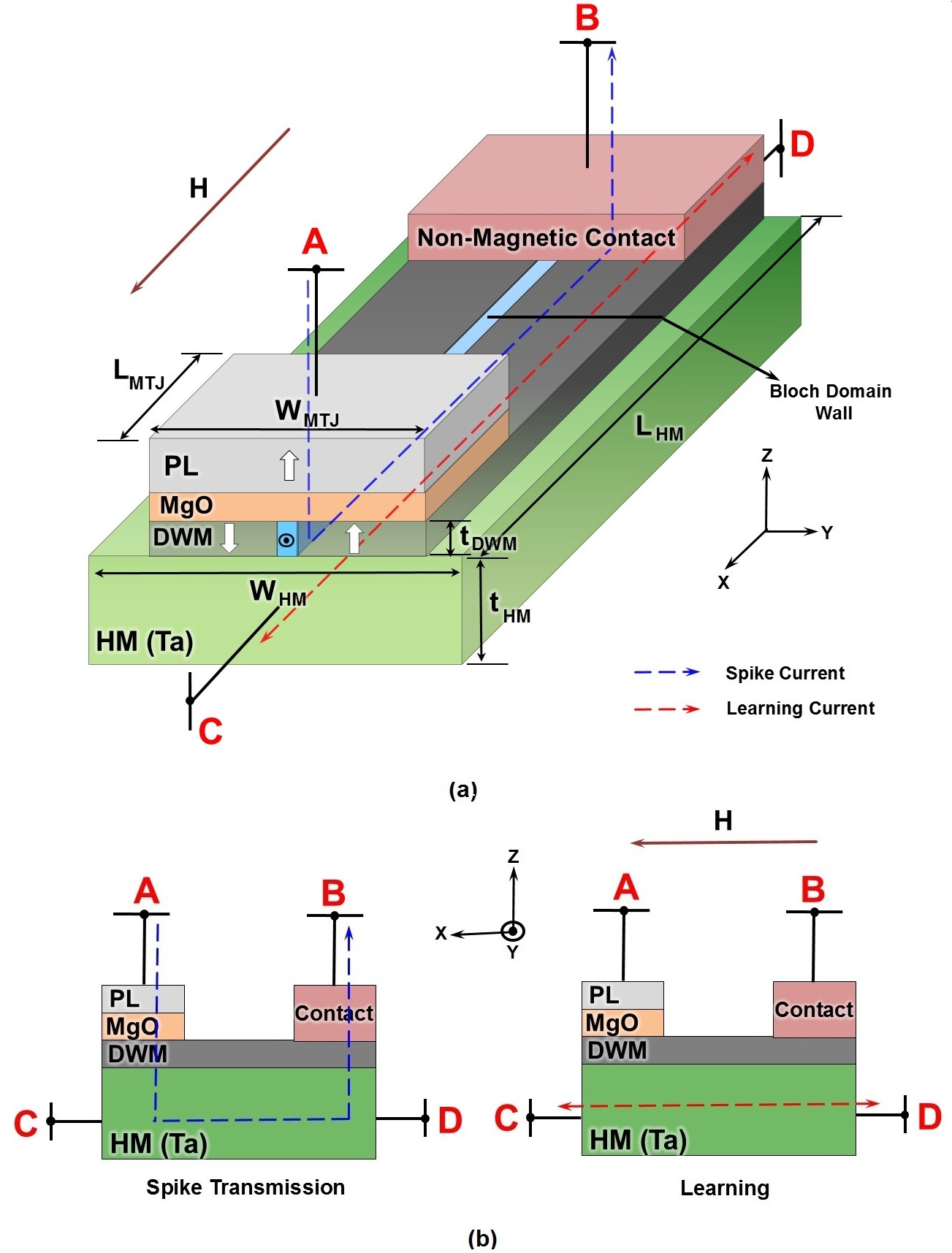}
\caption{{\scriptsize (a) Proposed synaptic device structure, (b) Side-view of the device with spike transmission and learning current paths}}
\label{fig1}
\end{figure}

The proposed four terminal synaptic device structure is shown in Fig.~\ref{fig1}(a). It consists of a magnetic heterostructure where a magnetic material with Perpendicular Magnetic Anisotropy (PMA) is in contact with a non-magnetic HM with high spin-orbit coupling. The magnetic material is a DWM where the domain wall is longitudinal, running parallel to the length of the DWM. The DWM is also part of a Magnetic Tunneling Junction (MTJ) structure where a tunneling oxide barrier (MgO) separates the DWM from the Pinned Layer (PL). Fig.~\ref{fig1}(b) depicts the side-view of the device structure. The spike current from the pre-neuron passes between terminals A and B through the MTJ structure. The learning current required to program the synapses flows through the HM between terminals C and D to implement STDP learning. An in-plane magnetic field $H$ is also applied during the learning stage. The spike transmission and learning operations are described next.

\begin{figure}[h]
\centering
\includegraphics[width=2.6in]{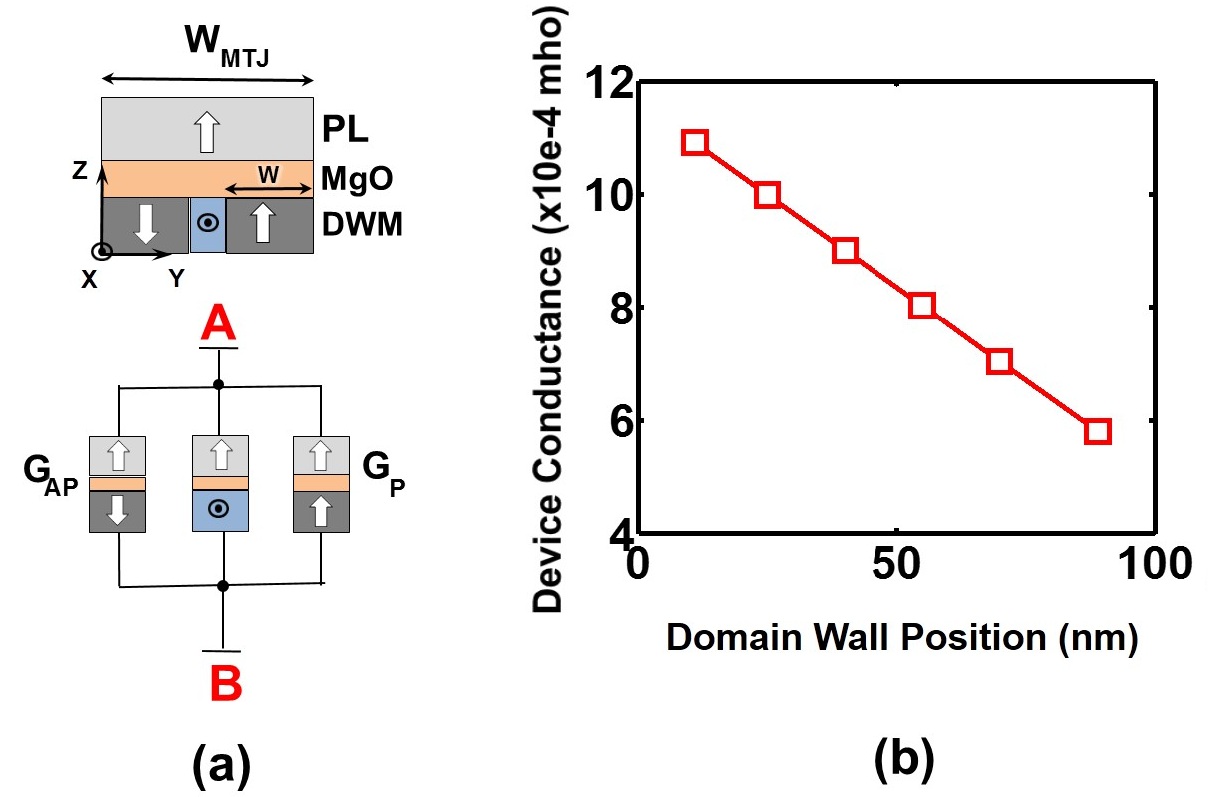}
\caption{{\scriptsize (a) Resistance model of the proposed synaptic device structure, (b) Variation of device conductance as a function of the domain wall position}}
\label{fig2}
\end{figure}

The resistance model of the device is shown in Fig.~\ref{fig2}(a). Considering the total width of the MTJ to be $W_{MTJ}$ and the width of the ferromagnetic domain whose magnetization is parallel to the PL to be $w$, the equivalent conductance of the device can be expressed as, 
\begin{equation}
     G_{dev}=G_{AP,max}(1-\frac{w}{W_{MTJ}})+G_{P,max}(\frac{w}{W_{MTJ}})+G_{DW}
\end{equation}
Here, $G_{AP,max}$ ($G_{P,max}$) represents the conductance of the device when the entire DWM magnetization is oriented anti-parallel (parallel) to the PL and $G_{DW}$ represents the conductance of the domain wall. Hence the device conductance varies linearly with the domain wall position as demonstrated in Fig.~\ref{fig2}(b). Non-equilibrium Green's function (NEGF) based transport simulation framework ~\cite{negf} was used to simulate the variation of the device conductance with domain wall position. The resistance of the DWM-HM heterostructure that lies in the path of the spike current between terminals A and B is negligible in comparison to the resistance of the tunneling oxide barrier. Hence, when a voltage spike from the pre-neuron is applied between terminals A and B, the device conductance will determine the strength of the spike current transmitted which can be modulated by programming the domain wall position. 

In order to implement STDP in the device, a current is passed between terminals C and D. When a programming current flows from terminal C to terminal D through the HM in the -x direction, spin-orbit torque leads to the accumulation of +y directed spin-polarized electrons at the HM-DWM interface. Negligible Dzyaloshinskii-Moriiya Interaction (DMI) and shape anisotropy due to the formation of the longitudinal domain wall leads to the formation of a Bloch wall in the sample ~\cite{sayeef}. The external in-plane magnetic field orients the magnetic moment of the domain wall along its direction. Thus the direction of the domain wall movement is determined by the cross-product of the accumulated spins at the HM-DWM interface and the direction of the applied magnetic field. For a magnetic field applied along the +x direction, application of current through the HM in the -x direction results in domain wall motion in the -y direction so that +z magnetic domain in the ferromagnet starts to dominate. It is worth noting here that conventional bulk spin-transfer torque does not contribute to the domain wall movement due to the formation of a longitudinal wall.

The magnetization dynamics of the ferromagnet can be described by solving Landau-Lifshitz-Gilbert equation with additional term to account for the spin momentum torque generated by the accumulated spin current at the HM-DWM interface ~\cite{slonc},
\begin{equation}
      \frac {d\widehat {\textbf {m}}} {dt} = -\gamma(\widehat {\textbf {m}} \times \textbf {H}_{eff})+ \alpha (\widehat {\textbf {m}} \times \frac {d\widehat {\textbf {m}}} {dt})+\beta (\widehat {\textbf {m}} \times \widehat {\textbf {m}}_P \times \widehat {\textbf {m}})
\end{equation}
where $\widehat {\textbf {m}}$ is the unit vector of DWM magnetization at each grid point, $\gamma= \frac {2 \mu _B \mu_0} {\hbar}$ is the gyromagnetic ratio for electron, $\alpha$ is Gilbert \textquoteright s damping ratio, $\textbf{H}_{eff}$ is the effective magnetic field, $\beta=\frac{\hbar P \theta J}{2 \mu_0 e t M_s}$ ( $\hbar$ is Planck’s constant, $P$ is polarization of the PL, $J$ is input charge current density, $\theta$ is spin-orbit torque efficiency, $\mu_0$ is permeability of vacuum, $e$ is electronic charge, $t$ is FL thickness and $M_s$ is saturation magnetization) and $\widehat {\textbf {m}}_P$ is direction of input spin current. Micromagnetic simulations were performed in $mumax^3$, a GPU-accelerated micromagnetic simulation program ~\cite{mumax}. The simulation parameters are given in Table I and was used for the rest of this work, unless otherwise stated.

\begin{table}[h]
\renewcommand{\arraystretch}{1.3}
\caption{Simulation Parameters}
\label{table_1}
\centering
\begin{tabular}{c c}
\hline 
\bfseries {\scriptsize Parameters} & \bfseries {\scriptsize Value}\\
\hline
{\scriptsize Ferromagnet Dimensions} &  {\scriptsize $200 \times 100 \times 1 nm^3$} \\
{\scriptsize Grid Size} & {\scriptsize $ 2 \times 2 \times 1 nm^3$} \\
{\scriptsize Heavy Metal Dimensions} & {\scriptsize $ 200 \times 1000 \times 10 nm^3$} \\
{\scriptsize Domain Wall Width} & {\scriptsize $ 22 nm$} \\
{\scriptsize MTJ (PL) Dimensions} & {\scriptsize $ 120 \times 100 \times 1 nm^3$ }\\
{\scriptsize Saturation Magnetization} & {\scriptsize 800 $KA/m$} \\
{\scriptsize Spin Orbit Torque Efficiency} & {\scriptsize 0.08} \\
{\scriptsize Gilbert Damping Factor} & {\scriptsize 0.024} \\
{\scriptsize MgO Thickness} & {\scriptsize 1.2$nm$} \\
{\scriptsize Exchange Correlation Constant} & {\scriptsize $3 \times 10^{-11} J/m$} \\
{\scriptsize Perpendicular Magnetic Anisotropy} & {\scriptsize $6 \times 10^{5} J/m^{-3}$} \\
{\scriptsize Magnetic Field} & {\scriptsize 10 $G$} \\
\hline
\end{tabular}
\end{table}

\begin{figure}[b]
\centering
\includegraphics[width=2.7in]{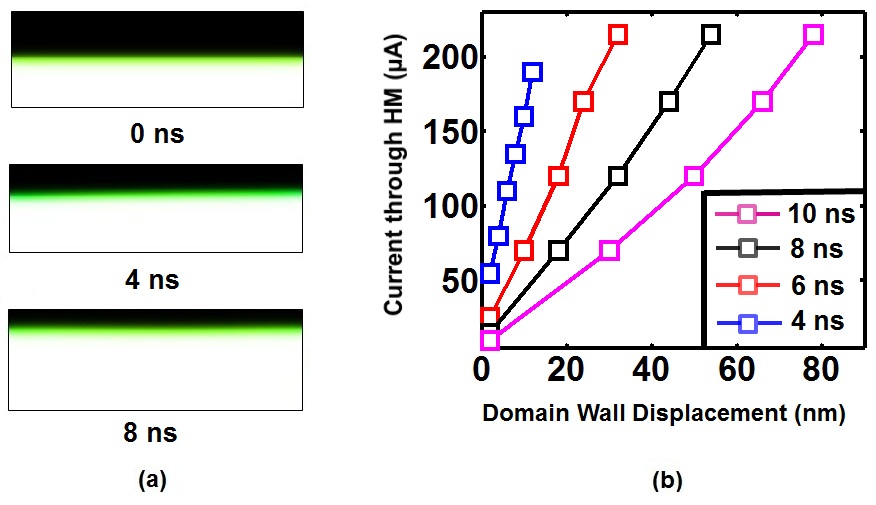}
\caption{{\scriptsize (a) Position of the domain wall in Ta-CoFeB heterostructure with CoFeB dimensions of $ 600 \times 200 \times 1 nm^3$ as a function of time, due to the application of a current density of $J=3.5 \times 10^6 A/cm^2$, (b) Variation of domain wall displacement with programming current through HM for different programming time durations}}
\label{fig3}
\end{figure}

The simulation framework was calibrated with experimental results demonstrated in ~\cite{sayeef} for Ta (HM) - CoFeB (DWM) heterostructure. Fig.~\ref{fig3}(a) depicts the position of the domain wall in the sample with CoFeB dimensions of $ 600 \times 200 \times 1 nm^3$ as a function of time, due to the application of a current density of $J=3.5 \times 10^6 A/cm^2$. The domain wall positions predicted by our simulation framework are in good agreement with the experimental results ~\cite{sayeef}.

For a given duration of the programming current, the domain wall displacement increases linearly with the magnitude of the current density. Fig.~\ref{fig3}(b) illustrates the linear increase of the domain wall displacement with programming current amplitude for different time durations. Since the device conductance is also a linear function of the domain wall position, the programming current follows a linear relationship with conductance change in the device. Reversing the direction of programming current or the direction of the magnetic field causes the domain wall to move in opposite direction. This enables us to implement STDP in the device as discussed in the following section.

\begin{figure}[h]
\centering
\includegraphics[width=2.5in]{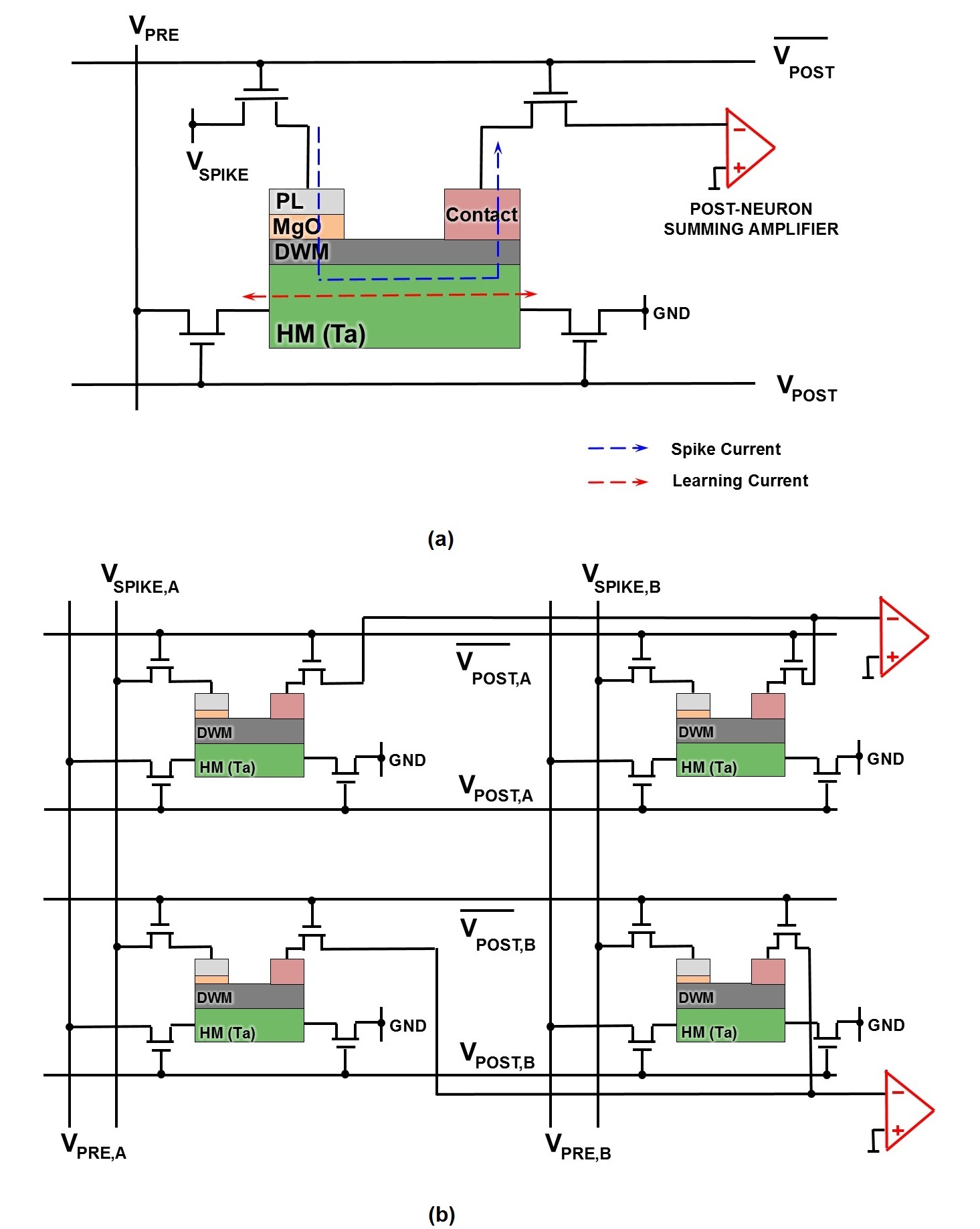}
\caption{{\scriptsize (a)Synaptic device with access transistors for separate spike transmission and learning current paths, (b) Possible arrangement of synapses in an array}}
\label{fig4}
\end{figure}

Fig.~\ref{fig4}(a) shows the proposed synaptic device with access transistors to decouple the spike transmission and learning current paths. A possible arrangement of the synapses in an array connecting the pre-neurons and post-neurons is depicted in Fig.~\ref{fig4}(b). When the pre-neuron spikes, the spike is transmitted using the signal $V_{SPIKE}$ through the MTJ structure. As long as the post-neuron does not spike, the spike transmission current path remains activated. Assuming that the resistance offered by the access transistors in the spike transmission current path is small in comparison to the resistance of the device, the spike voltage will be modulated by the device conductance and the weighted spike current will be transmitted to the post-neuron summing amplifier. As soon as the pre-neuron spikes, it also applies an appropriate programming voltage $V_{PRE}$ which extends over the time window to be used for learning. When the post-neuron spikes, the $V_{POST}$ signal gets activated. The $V_{POST}$ signal is a short pulse of duration of a few ns that essentially samples the appropriate amount of programming current from the $V_{PRE}$ signal. The spike transmission path gets de-activated and the appropriate programming current corresponding to the time delay between the pre-neuron and post-neuron spikes passes through the HM to move the domain wall to the appropriate location. Assuming that the magnetic field required for learning is generated by a local current carrying wire, the current through the wire can be turned on only when the post-neuron spikes. Hence spin-orbit torque is the underlying physical phenomena involved in the learning process as conventional bulk spin-transfer torque will not have any effect on the longitudinal domain wall. It is worth noting here that although some amount of spike current will flow through the HM, the magnitude of the spike current can be kept lower than the threshold current density required for domain wall movement to avoid any change in synaptic conductance.

\begin{figure}[h]
\centering
\includegraphics[width=2.5in]{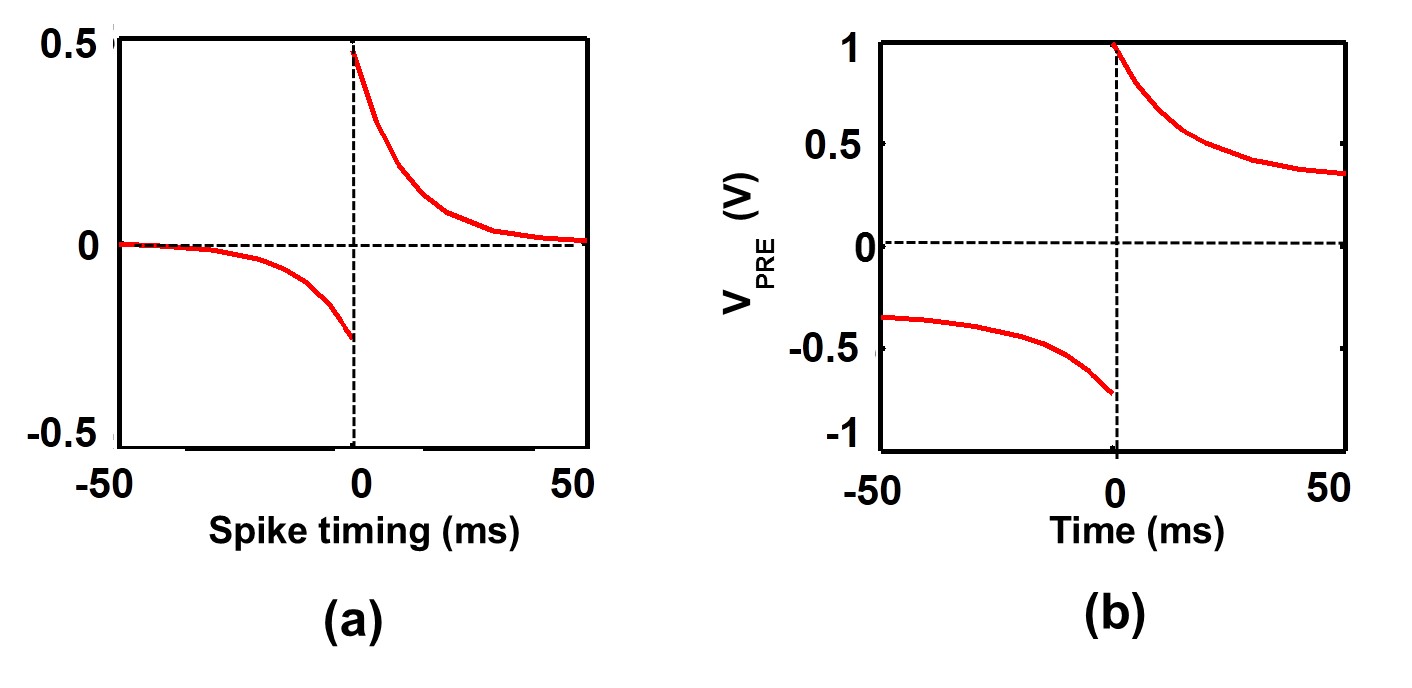}
\caption{{\scriptsize (a) Relative conductance change of proposed device as a function of spike timing, (c) $V_{PRE}$ programming signal as a function of time}}
\label{fig5}
\end{figure}

Fig.~\ref{fig5}(a) depicts the STDP characteristics implemented in our device which are in accordance to the characteristics measured in rat hippocampal glutamatergic synapses by Bi and Poo ~\cite{bipoo}. To account for learning during the negative timing window, the post-neuron programming signal $V_{POST}$ can be applied with a delay corresponding to the duration of the negative time window. For this work, the $V_{POST}$ signal was taken to be of duration $10 ns$. Once the $V_{POST}$ signal is activated, programming current flows through the device in case the $V_{PRE}$ signal remains active. Simulation of the programming circuit with access transistors was done using a commercial $45nm$ transistor model. The pre-neuron signal $V_{PRE}$ required to achieve the desired STDP characteristics is shown in Fig.~\ref{fig5}(b). For a time duration of $10ns$, the amount of programming current required to switch the DWM from the completely parallel to the anti-parallel state or vice-versa was found to be $\sim 200 \mu A$. Assuming that this current flows from a $1V$ supply, the corresponding energy consumption is $\sim 2 pJ ( V \times I \times t$). The desired programming current was achieved by appropriately sizing the access transistors for learning.

The major contributions of this work over state-of-the-art approaches can be summarized as follows. We propose a device structure comprising of decoupled spike transmission and programming current paths. While the learning current flows mainly through the HM and is responsible for generating STDP, the spike current gets modulated by the MTJ conductance, i.e. the strength of the synapse. 

The synaptic programming scheme is also simple and intuitive and has a direct correspondence to the learning scheme to be implemented. In most of the proposed programmable resistive synapses ~\cite{pcm1,pcm2,agsi,chalco} significant amount of programming voltage is applied across the device during the entire time duration of the learning window for the pre-neuron/ post-neuron leading to a large amount of redundant power consumption. Additionally they are characterized by relatively high programming threshold voltages $\sim$ a few volts. Our proposed programming scheme leads to current flow through the synapse only for a small time duration ($\sim$ a few $ns$) in case the post-neuron spikes before/ after the pre-neuron during the learning time window. Low programming currents are also required to modulate the device conductance due to high spin injection efficiency of spin-orbit torque. 

Ultra-low energy consumption of the order of $2pJ$ per synaptic event and the possibility of arranging such DWM-HM heterostructures in a crossbar fashion demonstrate the potential of this device as a possible candidate for an \textquoteleft electronic\textquoteright \ synapse in brain-inspired computing platforms.

\begin{acknowledgments}
The work was supported in part by, Center for Spintronic Materials, Interfaces, and Novel Architectures (C-SPIN), a MARCO and DARPA sponsored StarNet center, by the Semiconductor Research Corporation, the National Science Foundation, and by the National Security Science and Engineering Faculty Fellowship.
\end{acknowledgments}

\end{document}